# Molecular status of non-$q\bar{q}$ mesons.

Gerasyuta S.M.*, Kochkin V.I.

Department of Theoretical Physics, St. Petersburg State University, 198904, St. Petersburg, Russia.

Abstract.

The relativistic four-quark equations are found in the framework of the dispersion relation technique. The behavior of the low-energy four-particle amplitude is determined by its leading singularities in the pair invariant masses. The approximate solutions of these equations using the method based on the extraction of leading singularities of the amplitudes are obtained. The calculations of non-$q\bar{q}$ meson amplitudes estimate the contributions of three subamplitudes: four-quark amplitude, glueball amplitude and hadronic molecule amplitudes. The main contributions to non-$q\bar{q}$ meson amplitude are determined by the four-quark state and glueball. The contribution of hadronic molecule subamplitudes is only 10 % of non-$q\bar{q}$ meson amplitude.

---

\* Present address: Department of Physics, LTA, Institutski Per. 5, St. Petersburg 194021, Russia

## I. Introduction

QCD, the dynamically theory of strong interactions, appears to predict the existence of hadrons beyond those in the simple quark model, namely glueballs, hybrids and $q\bar{q}q\bar{q}$ states. Confirmation of such states would give information of the role of "dynamical" color in low energy QCD. The spectroscopy of low-mass states can be accounted by QCD - inspired models [1-9].

Phenomenological models are expected to provide a scheme for taking into account vacuum effects and other effects of a nonperturbative character in a zero-order approximation. Investigations revealed that the spectroscopy of light hadrons is one of the main sources of information about nonperturbative QCD effects.

In the series of papers [10 - 14] a practical treatment of relativistic three-hadron systems have been developed. The physics of three-hadron system is usefully described in terms of the pairwise interactions among the three particles. The theory is based on the two principles of unitarity and analiticity, as applied to the two-body subenergy channels. The linear integral equations in a single variable are obtained for the isobar amplitudes. Instead of the quadrature methods of obtaining solution the set of suitable functions is identified and used as basis set for the expansion of the desired solutions. By this means the coupled integral equations are solved in terms of simple algebra.

In paper [15] the Faddeev equations are represented in form of the dispersion relation over two-body subenergy. The behavior of the low-energy three-particle amplitude is determined by its leading singularities in the pair invariant masses. Then the purpose was to extract the singular part of the amplitude. The suggested method of approximate solution of the Faddeev equations was verified on the example of simplest low-energy NN-interaction (the S-wave scattering length approximation). In this approximation the wavefunction was found which describes well the form factor of tritium/helium-3 at small $q^2$ values.

In [16, 17] a relativistic generalization of the three-body Faddeev equations was obtained in the form of dispersion relations in the pair energy of two interacting particles. The mass spectrum of S-wave baryons including u, d and s quarks was calculated by a method based on isolating of the leading singularities in the amplitude. We searched for the approximate solution of integral three-quark equations by taking into account two-particle and triangle singularities, all the weaker ones being neglected. If we considered the approximation, which corresponds to taking into account two-body and triangle singularities, and define all the smooth functions of the subenergy variables (as compared with the singular part of the amplitude) in the middle point of physical region of Dalitz-plot, then the problem reduces to one of solving simple algebraic system equations.

In the recent papers [18, 19] the relativistic generalization of the four-body Faddeev-Yakubovsky type equations are represented in the form of dispersion relations over the two-body subenergy. The four-quark amplitudes for low-lying cryptoexotic mesons were calculated under the condition that flavor SU(3) symmetry holds. It should be noted, that even in this approximation, the calculated masses of lowest cryptoexotic mesons agree well with experimental data [20] and with the results obtained in the other models [21-26].

In the present paper the relativistic four-quark equations (like Faddeev-Yakubovsky equations) are constructed in the form of the dispersion relation over the two-body subenergy.

We calculated the masses of the non-$q\bar{q}$ mesons using method based on the extraction of leading singularities of the non-$q\bar q$ meson amplitudes. The main results of this paper is the calculation of non-$q\bar q$ meson amplitudes, which contain the contributions of three subamplitudes: four-quark amplitude, glueball amplitude and hadronic molecule amplitudes. The contribution of hadronic molecule amplitudes is less than 10 % of non-$q\bar q$ meson amplitude.

In Section II the relativistic four-quark equations are constructed in the form of the dispersion relation over the two-body subenergy. The approximate solutions of these equations using the method based on the extraction of leading singularities of the amplitude are obtained. We used the classification of singularities, which was proposed in paper [27]. One considers the approximation, which corresponds to the single interaction of all four-particles (two-particle, triangle and four-particle singularities).

Section III is devoted to the calculation of three subamplitudes: four-quark amplitude, glueball amplitude and hadronic molecule amplitudes (Table I).

In the Conclusion the status of the considered model is discussed.

In the Appendix A the quark-antiquark vertex functions and the phase spaces for the hybrid mesons are given respectively (Tables II, III).

In the Appendix B we search the integration contours of functions $J_1$, $J_2$, $J_3$, which are determined by the interaction of the four quarks.

II. Quark amplitudes of non-$q\bar q$ mesons

We derived the relativistic four-quark equations in the framework of the dispersion relation technique. For the sake of simplicity one considers the case of the $SU(3)_f$ - symmetry, that the masses of all quarks are equal. We consider only planar diagrams, the other diagrams due to the rules of $1/N_c$ expansion [28, 29] are neglected. The correct equations for the amplitude are obtained at the account of all possible subamplitudes. It corresponds to the division complete system into subsystems from the smaller number of particles. Then one should represent four-particle amplitude as a sum of six subamplitudes: $A = A_{12} + A_{13} + A_{14} + A_{23} + A_{24} + A_{34}$. In our case all particles are identical, therefore we need to consider only one group of diagrams and the amplitude corresponding to them, for example $A_{12}$. We must take into account each sequence of the inclusion of interaction. For instance, the process beginning with interaction of the particles 1 and 2 can proceed by the three ways: particle 3 and 4 consistently join a chosen pair, or begin to interact among themselves, and each of the three ways of the connection there should correspond to their own amplitudes. Therefore the diagrams corresponding to amplitude $A_{12}$ are divided in two group (if the particles are different in three groups). The graphic equations are shown Fig.1. The coefficients are determined by the permutation of quarks [30, 31].

To represent the four-quark $A_1(s, s_{12}, s_{123})$, glueball $A_2(s, s_{12}, s_{34})$ and hadronic molecule amplitudes $A_p(s, s_{12}, s_{34})$ (p=3, 4, 5, 6, 7) in form of the dispersion relation it is necessary to define the amplitude of two-quark interaction $b_j(s_{ik})$. The quark amplitudes $q\bar q \to q\bar q$ are calculated in the framework of the dispersion N/D method with the input four-fermion interaction with quantum numbers of the gluon. We use the results of the bootstrap quark model [32] and write down the pair quarks amplitude in the form:

$$b_j(s_{ik}) = \frac{G_j^2(s_{ik})}{1-B_j(s_{ik})},\qquad(1)$$

$$B_j(s_{ik}) = \int_{4m^2}^{\Lambda}\frac{ds'_{ik}}{\pi}\frac{\rho_j(s'_{ik})G_j^2(s'_{ik})}{s'_{ik}-s_{ik}}.\qquad(2)$$

Here $s_{ik}$ is the two-particle subenergy squared, $s_{ijk}$ corresponds to the energy squared of particles $i$, $j$, $k$ and $s$ is the system total energy squared. $G_j(s_{ik})$ is the quark-antiquark vertex function. (Table II) $B_j(s_{ik})$, $\rho_j(s_{ik})$ are the Chew-Mandelstam function and the phase space respectively. They are given in the Appendix A. We introduced the cut-off parameter $\Lambda$. There j=1 corresponds to pair of quarks $q\bar{q}$ with $J^{PC} = 0^{++}$, $1^{++}$, $2^{++}$, $0^{-+}$, $1^{--}$ (color singlet $SU(3)_c$) and j=2 defines the quark pair with $J^{PC} = 1^{--}$ in color channel $8_c$ (constituent gluon). In the case in question the interacting quarks do not produce bound state, then the integration in (3) - (5) is carried out from the threshold $4m^2$ to the cut-off $\Lambda$. The integral equation systems, corresponding to Fig. 1, have the following form:

$$A_1(s,s_{12},s_{123}) = \frac{\lambda_1 B_1(s_{12})}{1-B_1(s_{12})} + \frac{2G_1(s_{12})}{1-B_1(s_{12})}[\hat{J}_1 A_1(s,s'_{13},s_{123}) + \hat{J}_3 A_2(s,s'_{13},s'_{24}) +$$
$$+\sum_{p=3}^{7}\hat{J}_{3p}A_p(s,s'_{13},s'_{24})]\qquad(3)$$

$$A_2(s,s_{12},s_{34}) = \frac{\lambda_2 B_2(s_{12})B_2(s_{34})}{[1-B_2(s_{12})][1-B_2(s_{34})]} + \frac{4G_2(s_{12})G_2(s_{34})}{[1-B_2(s_{12})][1-B_2(s_{34})]}\hat{J}_2 A_1(s,s'_{13},s'_{134}),\qquad(4)$$

$$A_p(s,s_{12},s_{34}) = \frac{\lambda_p B_p(s_{12})B_p(s_{34})}{[1-B_p(s_{12})][1-B_p(s_{34})]} + \frac{4G_p(s_{12})G_p(s_{34})}{[1-B_p(s_{12})][1-B_p(s_{34})]}\hat{J}_{2p} A_1(s,s'_{13},s'_{134}),\qquad(5)$$

here $G_p(s_{ik}) \equiv G_1(s_{ik},J_p^{PC})$, $B_p(s_{ik}) \equiv B_1(s_{ik},J_p^{PC})$, p=3, 4, 5, 6, 7 correspond to $J^{PC} = 0^{++}$, $1^{++}$, $2^{++}$, $0^{-+}$, $1^{--}$ and $\lambda_i$ are the current constants. There we introduce the integral operators:

$$\hat{J}_1(s_{12},s'_{13}) = \int_{4m^2}^{\Lambda}\frac{ds'_{12}}{\pi}\frac{\rho_1(s'_{12})\cdot G_1(s'_{12})}{s'_{12}-s_{12}}\int_{-1}^{+1}\frac{dz_1}{2},\qquad(6)$$

$$\hat{J}_2(s_{12},s_{34},s'_{13},s'_{134}) = \int_{4m^2}^{\Lambda}\frac{ds'_{12}}{\pi}\frac{\rho_2(s'_{12})\cdot G_2(s'_{12})}{s'_{12}-s_{12}}\int_{4m^2}^{\Lambda}\frac{ds'_{34}}{\pi}\frac{\rho_2(s'_{34})\cdot G_2(s'_{34})}{s'_{34}-s_{34}}\int_{-1}^{+1}\frac{dz_3}{2}\int_{-1}^{+1}\frac{dz_4}{2},\qquad(7)$$

$$\hat{J}_3(s'_{12}, s_{123}, s'_{13}, s'_{24}) = \frac{1}{4\pi} \int\limits_{4m^2}^{\Lambda} \frac{ds'_{12}}{\pi} \frac{\varrho_1(s'_{12}) \cdot G_1(s'_{12})}{s'_{12} - s_{12}} \int\limits_{-1}^{+1} \frac{dz_1}{2} \int\limits_{-1}^{+1} dz \int\limits_{z_2^-}^{z_2^+} dz_2 \times$$
$$\times \frac{1}{\sqrt{1 - z^2 - z_1^2 - z_2^2 + 2zz_1z_2}},$$
(8)

$$\hat{J}_{2p} = \hat{J}_2, \quad \hat{J}_{3p} = \hat{J}_3 \tag{9}$$

at $G_2(s_{ik}) \to G_p(s_{ik}) = G_1(s_{ik}, J^{PC})$, $\varrho_2 \to \varrho_p \equiv \varrho_1(s_{ik}, J_p^{PC})$, p=3, 4, 5, 6, 7 correspond to $J^{PC} = 0^{++}, 1^{++}, 2^{++}, 0^{-+}, 1^{--}$. There $m$ is the quark mass.

Hereafter we suggest that is some unknown (not large) contribution from small distances which might be taken into account with the help of cut-off procedure. In the (6)-(9) we choose the "hard" cutting, but we can use also the "soft" cutting, for instance $G_i(s_{ik}) = G_i \exp\left(-(s_{ik} - 4m^2)^2/\Lambda^2\right)$ and do not change essentially the calculated mass spectrum. In the equations (6) and (8) $z_1$ is the cosine of the angle between the relative momentum of the particles 1 and 2 in the intermediate state and that of the particle 3 in the final state, which is taken in the c.m. of particles 1 and 2. In the equation (8) $z$ is the cosine of the angle between the momentum of the particles 3 and 4 in the final state, which is taken in the c.m. of particles 1 and 2. $z_2$ is the cosine of the angle between the relative momentum of particles 1 and 2 in the intermediate state and the momentum of the particle 4 in the final state, which is taken in the c.m. of particles 1 and 2. In the equation (7), (9) we have defined that: $z_3$ is the cosine of the angle between relative momentum of particles 1 and 2 in the intermediate state and that of the relative momentum of particles 3 and 4 in the intermediate state, which is taken in the c.m. of particles 1 and 2. $z_4$ is the cosine of the angle between the relative momentum of the particles 3 and 4 in the intermediate state and that of the momentum of the particle 1 in the intermediate state which is taken in the c.m. of particles 3, 4. Using (10) - (14) we can pass from the integration over the cosines of the angles to the integration over the subenergies.

$$s'_{13} = 2m^2 + \frac{s_{123} - s'_{12} - m^2}{2} + \frac{z_1}{2}\sqrt{\frac{s'_{12} - 4m^2}{s'_{12}}[(s_{123} - s'_{12} - m^2)^2 - 4s'_{12}m^2]}, \tag{10}$$

$$s'_{24} = 2m^2 + \frac{s'_{124} - s'_{12} - m^2}{2} + \frac{z_2}{2}\sqrt{\frac{s'_{12} - 4m^2}{s'_{12}}[(s'_{124} - s'_{12} - m^2)^2 - 4s'_{12}m^2]}, \tag{11}$$

$$z = \frac{2s'_{12}(s + s'_{12} - s_{123} - s'_{124}) - (s_{123} - s'_{12} - m^2)(s'_{124} - s'_{12} - m^2)}{\sqrt{[(s_{123} - s'_{12} - m^2)^2 - 4m^2 s'_{12}][(s'_{124} - s'_{12} - m^2)^2 - 4m^2 s'_{12}]}}, \tag{12}$$

$$s'_{134} = m^2 + s'_{34} + \frac{s - s'_{12} - s'_{34}}{2} + \frac{z_3}{2}\sqrt{\frac{s'_{12} - 4m^2}{s'_{12}}[(s - s'_{12} - s'_{34})^2 - 4s'_{12}s'_{34}]}, \tag{13}$$

$$s'_{13} = 2m^2 + \frac{s'_{134} - s'_{34} - m^2}{2} + \frac{z_4}{2}\sqrt{\frac{s'_{34} - 4m^2}{s'_{34}}\left[(s'_{134} - s'_{34} - m^2)^2 - 4m^2 s'_{34}\right]}. \tag{14}$$

Let us extract two-particle singularities in the amplitudes $A_1(s, s_{12}, s_{123})$, $A_2(s, s_{12}, s_{34})$ and $A_p(s, s_{12}, s_{34})$ (p=3, 4, 5, 6, 7):

$$A_1(s, s_{12}, s_{123}) = \frac{\alpha_1(s, s_{12}, s_{123}) B_1(s_{12})}{1 - B_1(s_{12})}, \tag{15}$$

$$A_2(s, s_{12}, s_{34}) = \frac{\alpha_2(s, s_{12}, s_{34}) B_2(s_{12}) B_2(s_{34})}{[1 - B_2(s_{12})][1 - B_2(s_{34})]}, \tag{16}$$

$$A_p(s, s_{12}, s_{34}) = \frac{\alpha_p(s, s_{12}, s_{34}) B_p(s_{12}) B_p(s_{34})}{[1 - B_p(s_{12})][1 - B_p(s_{34})]}. \tag{17}$$

In the amplitude $A_1(s, s_{12}, s_{123})$ we do not extract three-particle singularity, because it is weaker than two-particle and taking into account in the function $\alpha_1(s, s_{12}, s_{123})$.

We used the classification of singularities, which was proposed in paper [27]. The construction of approximate solution of the (15) - (17) is based on the extraction of the leading singularities of the amplitudes. The main singularities in $s_{ik} \approx 4m^2$ are from pair rescattering of the particles i and k. First of all there are threshold square root singularities. Also possible are pole singularities which correspond to the bound states. They are situated on the first sheet of complex $s_{ik}$ plane in case of real bound state and on the second sheet in case of virtual bound state. The diagrams Fig.1 apart from two-particle singularities have their specific triangular singularities and the singularities correspond to the interaction of four particles. Such classification allows us to search the approximate solution of (15) - (17) by taking into account some definite number of leading singularities and neglecting all the weaker ones. We consider the approximation, which corresponds to the single interaction of all four particles (two-particle, triangle and four-particle singularities). The functions $\alpha_1(s, s_{12}, s_{123})$, $\alpha_2(s, s_{12}, s_{34})$, $\alpha_p(s, s_{12}, s_{34})$ are the smooth functions of $s_{ik}$, $s_{ijk}$ as compared with the singular part of the amplitudes, hence they can be expanded in a series in the singularity point and only the first term of this series should be employed further. Using this classification one define the reduced amplitudes $\alpha_1(s, s_{12}, s_{123})$, $\alpha_2(s, s_{12}, s_{34})$ and $\alpha_p(s, s_{12}, s_{34})$ as well as the B-functions in the middle point of the physical region of Dalitz-plot at the point $s_0$:

$$s_0 = (s + 8m^2)/6, \quad s_{123} = 3s_0 - 3m^2. \tag{18}$$

Such a choice of points $s_0$ allows as to replace the integral equations (3) - (5) by the algebraic equations (19) - (21) respectively:

$$a_1 = \lambda_1 + 2\left[ a_1 J_1(s_0) + a_2 J_3(s_0) \frac{B_2^2(s_0)}{B_1(s_0)} + \sum_{p=3}^{7} a_p J_{3p}(s_0) \frac{B_p^2(s_0)}{B_1(s_0)} \right], \tag{19}$$

$$a_2 = \lambda_2 + 4 a_1 J_2(s_0) \frac{B_1(s_0)}{B_2^2(s_0)}, \tag{20}$$

$$a_p = \lambda_p + 4a_1 J_{2p}(s_0) \frac{B_1(s_0)}{B_p^2(s_0)}, \tag{21}$$

p=3, 4, 5, 6, 7.

One obtained, that the other choice of point $s_0$ do not change essentially the contributions of $\alpha_1$, $\alpha_2$ and $\alpha_p$.

Here we introduce following functions:

$$J_1(s_0) = G_1^2 \int\limits_{4m^2}^{\Lambda} \frac{ds'_{12}}{\pi} \frac{\varrho_1(s'_{12})}{s'_{12}-s_0} \int\limits_{-1}^{+1} \frac{dz_1}{2} \frac{1}{1-B_1(s'_{13})}, \tag{22}$$

$$J_2(s_0) = G_2^4 \int\limits_{4m^2}^{\Lambda} \frac{ds'_{12}}{\pi} \frac{\varrho_2(s'_{12})}{s'_{12}-s_0} \int\limits_{4m^2}^{\Lambda} \frac{ds'_{34}}{\pi} \frac{\varrho_2(s'_{34})}{s'_{34}-s_0} \int\limits_{-1}^{+1} \frac{dz_3}{2} \int\limits_{-1}^{+1} \frac{dz_4}{2} \frac{1}{1-B_1(s'_{13})}, \tag{23}$$

$$J_3(s_0) = G_1^2 \frac{1-B_1(s_0, \Lambda)}{1-B_1(s_0, \tilde{\Lambda})} \frac{1}{4\pi} \int\limits_{4m^2}^{\tilde{\Lambda}} \frac{ds'_{12}}{\pi} \frac{\varrho_1(s'_{12})}{s'_{12}-s_0} \int\limits_{-1}^{+1} \frac{dz_1}{2} \int\limits_{-1}^{+1} dz \int\limits_{z_2^-}^{z_2^+} dz_2 \times$$

$$\times \frac{1}{\sqrt{1-z^2-z_1^2-z_2^2+2zz_1z_2}} \frac{1}{[1-B_2(s'_{13})][1-B_2(s'_{24})]}, \tag{24}$$

$J_{2p}(s_0) = J_2(s_0)$, $J_{3p}(s_0) = J_3(s_0)$, at $B_2(s_{ik}) \to B_p(s_{ik}) \equiv B_1(s_{ik}, J_p^{PC})$, p=3, 4, 5, 6, 7 correspond to $J^{PC} = 0^{++}$, $1^{++}$, $2^{++}$, $0^{-+}$, $1^{--}$. In our approximation the vertex functions (Table II) are constants. As the integration region the physical region of the reaction should be chosen, therefore $-1 \le z_i \le 1$ ( i=1,2,3,4 ). From these conditions we can define the regions of the integration over $s'_{13}$, $s'_{24}$, $s'_{134}$, $s'_{124}$. Let us consider the integration region over $s'_{124}$. For this purpose we use equation (12). This condition corresponds to $0 \le z^2 \le 1$. By consideration of these inequalities one can obtain:

$$s_{124}^{\pm} = s'_{12} + m^2 + \frac{(s-s_{123}-m^2)(s_{123}+s'_{12}-m^2)}{2s_{123}} \pm$$

$$\pm \frac{1}{2s_{123}} \sqrt{[(s_{123}-s'_{12}-m^2)^2 - 4m^2 s'_{12}][(s-s_{123}-m^2)^2 - 4m^2 s_{123}]} \tag{25}$$

We must take into account the upper restriction of the integration region over $s'_{12}$ in $J_3$:

$$\tilde{\Lambda} = \begin{cases} \Lambda, & \text{if } \Lambda \le (\sqrt{s_{123}}+m)^2 \\ (\sqrt{s_{123}}+m)^2, & \text{if } \Lambda > (\sqrt{s_{123}}+m)^2 \end{cases} \tag{26}$$

The integration contours of the functions $J_1, J_2, J_3$ are given in the Appendix B. The function $J_3$ takes into account the singularity, which corresponds to the simultaneous vanishing of all propagators in the four-particle diagram like those in Fig.1. In the case in question the functions $\alpha_i(s)$ are determined as:

$$\alpha_i(s) = F_i(s, \lambda_i) / \Delta(s). \qquad (27)$$

There $\Delta(s)$ is the determinant:

$$\Delta(s) = 1 - 2J_1 - 8J_2 J_3 - 8\sum_{p=3}^{7} J_{2p} J_{3p}. \qquad (28)$$

The poles of the reduced amplitudes $\alpha_1$, $\alpha_2$, $\alpha_p$ for the lowest non-$q\bar{q}$ mesons with $J^{PC} = 0^{++}, 1^{++}, 2^{++}, 0^{-+}, 1^{--}$ correspond to the bound state and determine the masses of the non-$q\bar{q}$ mesons.

### III. Calculation results

In the considered calculation the quark masses $m$ is not fixed. In order to fix anyhow $m$, we assume $m = 438\,MeV$ ($m \geq \frac{1}{4} m_{f_2}(1710)$). The model in question have two parameters: cut-off parameter $\Lambda$ and gluon constant $g$, which can be determined by mean of fixing of lowest non-$q\bar{q}$ meson mass values ($J^{PC} = 0^{-+}, 2^{++}$). The calculation values of mass lowest non-$q\bar{q}$ meson are shown in the Table I. The main results of this model is the calculation of non-$q\bar{q}$ meson amplitudes, which estimate the contributions of three subamplitudes are given in the Table I. We can see that the main contributions correspond to the four-quark amplitude $A_1$ and the glueball amplitude $A_2$ (Fig. 1). The contributions of the hadronic molecule amplitudes is less than 10 % ones. These results do not depend essentially on the calculated values of mass lowest non-$q\bar{q}$ meson. We obtain that the contributions of hadronic molecules to the non-$q\bar{q}$ mesons is small as compared the contributions of the four-quark amplitude and glueball amplitude.

### IV. Conclusion

In the framework of approximate method of solution four-particle relativistic equations the mass spectrum of lowest non-$q\bar{q}$ mesons are calculated. We use the four-fermion interaction with quantum number of the gluon. In the bootstrap quark model [32] there is a bound state in the gluon channel with mass of the order 0.7 GeV. This bound state should be identified as a constituent gluon. The resulting quark interaction appeared to be effectively short-range. This interaction is determined mainly by the exchange in the gluon channel: the constituent gluon mass appeared to be not small. The value constituent gluon mass obtained in this model seems to be rather reasonable: just this mass value is required by hadron phenomenology [33-35].

We manage with the quarks as with real particles. However, in the soft region the quark diagrams should be treated as spectral integrals over quark mass with the spectral density $\rho(m^2)$: the integration over quark masses in the amplitudes puts away the quark singularities and introduces the hadron ones. One can believe that the approximation:

$$\rho(m^2) \Rightarrow \delta(m^2 - m_q^2), \qquad (29)$$

could be possible for the low-lying hadrons (here $m_q$ is the "mass" constituent quark). We hope the approach given by (29) is sufficiently good for the calculation of the low-lying non-$q\bar{q}$ mesons being carried out here. The problem of distribution over quark masses is important when one considers that the high excited states need special studies.

The decay width of non-$q\bar{q}$ mesons can be calculated in the framework of this model. The suggested approximate method allows to construct the four-quark amplitudes, including quarks of three flavours (u, d, s) and taking account only $q\bar{q}q\bar{q}$ states and glueball.

Acknowledgments.

The authors would like to thank T. Barnes, V.A. Franke and Yu.V. Novozhilov for useful discussions. S.M. Gerasyuta thanks St. Petersburg University for the hospitality where a part of this work was completed.

# APPENDIX A

The two-particle phase space for the equal quark masses is defined as:

$$\rho_1(s_{ik}, J^{PC}) = \left(\gamma(J^{PC})\frac{s_{ik}}{4m^2} + \beta(J^{PC})\right)\sqrt{\frac{s_{ik} - 4m^2}{s_{ik}}},$$

$$\rho_2(s_{ik}) = \rho_1(s_{ik}, 1^{--}),$$

$\rho_p(s_{ik}) = \rho_1(s_{ik}, J_p^{PC})$, p=3, 4, 5, 6, 7 correspond to $J^{PC} = 0^{++}$, $1^{++}$, $2^{++}$, $0^{-+}$, $1^{--}$.

The vertex functions are shown in Table II. The coefficients $\gamma(J^{PC})$ and $\beta(J^{PC})$ are given in Table III.

# APPENDIX B

The integration contour 1 (Fig. 2) corresponds to the connection $s_{123} < (\sqrt{s_{12}} - m)^2$, the contour 2 is defined by the connection $(\sqrt{s_{12}} - m)^2 < s_{123} < (\sqrt{s_{12}} + m)^2$. The point $s_{123} = (\sqrt{s_{12}} - m)^2$ is not singular, that the round of this point at $s_{123} + i\varepsilon$ and $s_{123} - i\varepsilon$ gives identical result. $s_{123} = (\sqrt{s_{12}} + m)^2$ is the singular point, but in our case the integration contour can not pass through this point that the region in consideration is situated below the production threshold of the four particles $s < 16m^2$. The similar situation for the integration over $s_{13}$ in the function $J_3$ is occurred. But the difference consists of the given integration region that is conducted between the complex conjugate points (contour 2 Fig. 2). In Fig. 2, 3b, 4 the dotted lines define the square root cut of the Chew-Mandelstam functions. They correspond to two-particles threshold and also three-particles threshold in Fig. 3(a). The integration contour 1 (Fig. 3(a)) is determined by $s < (\sqrt{s_{12}} - \sqrt{s_{34}})^2$, the contour 2 corresponds to the case $(\sqrt{s_{12}} - \sqrt{s_{34}})^2 < s < (\sqrt{s_{12}} + \sqrt{s_{34}})^2$. $s = (\sqrt{s_{12}} - \sqrt{s_{34}})^2$ is not singular point, that the round of this point at $s + i\varepsilon$ and $s - i\varepsilon$ gives identical results. The integration contour 1 (Fig. 3(b)) is determined by region $s < (\sqrt{s_{12}} - \sqrt{s_{34}})^2$ and $s_{134} < (\sqrt{s_{34}} - m)^2$, the integration contour 2 corresponds to $s < (\sqrt{s_{12}} - \sqrt{s_{34}})^2$ and $(\sqrt{s_{34}} - m)^2 \leq s_{134} < (\sqrt{s_{34}} + m)^2$. The contour 3 is defined by $(\sqrt{s_{12}} - \sqrt{s_{34}})^2 < s < (\sqrt{s_{12}} + \sqrt{s_{34}})^2$. Here the singular point would be $s_{134} = (\sqrt{s_{34}} + m)^2$. But in our case this point is not achievable, if one has the condition $s < 16m^2$. We have to consider the integration over $s_{24}$ in the function $J_3$. While $s_{124} < s_{12} + 5m^2$ the integration is conducted along the complex axis (the contour 1, Fig. 4). If we come to the point $s_{124} = s_{12} + 5m^2$, that the output into the square root cut of Chew-Mandelstam function (contour 2, Fig. 4) is occurred. In this case the part of the integration contour in nonphysical region is situated and the integration contour along the real axis is conducted. The other part of integration contour corresponds to physical regions. This part of integration contour along the complex axis is conducted. The suggested calculation shows that the contribution of the integration over the nonphysical region is small [36].

Table I. Low-lying non-$q\bar{q}$ meson masses and contributions of four-quark $A_1$, glueball $A_2$ and hadronic molecule $A_p$ (p=3, 4, 5, 6, 7) subamplitudes to the non-$q\bar{q}$ meson amplitude in %.

| $J^{PC}$ | Masses (MeV) | $A_1$ | $A_2$ | $A_3$ $J^{PC}=0^{++}$ | $A_4$ $J^{PC}=1^{++}$ | $A_5$ $J^{PC}=2^{++}$ | $A_6$ $J^{PC}=0^{-+}$ | $A_7$ $J^{PC}=1^{--}$ |
|---|---|---|---|---|---|---|---|---|
| $0^{++}$ | 1593 $f_0$(1500) | 54.66 | 37.82 | 1.59 | 1.32 | 0.72 | 3.08 | 0.81 |
| $1^{++}$ | 1615 $f_1$(1510) | 53.72 | 38.72 | 1.59 | 1.32 | 0.73 | 3.11 | 0.81 |
| $2^{++}$ | 1710 $f_2$(1710) | 46.20 | 45.82 | 1.58 | 1.39 | 0.76 | 3.40 | 0.85 |
| $0^{-+}$ | 1440 $\eta$(1440) | 62.55 | 30.80 | 1.47 | 1.17 | 0.64 | 2.66 | 0.71 |
| $1^{--}$ | 1697 ( - ) | 47.49 | 44.57 | 1.59 | 1.39 | 0.76 | 3.35 | 0.85 |

Parameters of model: cut-off parameter $\Lambda$ =16; gluon constant $g$ =0,2687; effective mass $m$ = 438 MeV. Experimental mass values of the non-$q\bar{q}$ mesons are given in parentheses [20].

Table II. Vertex functions

| $J^{PC}$ | $G_1^2$ |
|---|---|
| $0^{++}$ | - 8g/3 |
| $1^{++}$ | 4g/3 |
| $2^{++}$ | 4g/3 |
| $0^{-+}$ | $8g/3 - 4g(m_i + m_k)^2/(3s_{ik})$ |
| $1^{--}$ | 4g/3 |

The vertex functions $G_1$ correspond to color singlet states. $G_2^2 = 3g$ correspond to the constituent gluon. Here $g$ is the gluon constant. In the present paper the contribution of axial interaction to the state $J^{PC}=0^{-+}$ is taken into account.

Table III. Coefficient of Chew-Mandelstam functions.

| $J^{PC}$ | $\gamma(J^{PC})$ | $\beta(J^{PC})$ |
|---|---|---|
| $0^{++}$ | -1/2 | 1/2 |
| $1^{++}$ | 1/2 | 0 |
| $2^{++}$ | 3/10 | 1/5 |
| $0^{-+}$ | 1/2 | 0 |
| $1^{--}$ | 1/3 | 1/6 |

Figure captions.

Fig.1. Graphic representation of the equations for the four-quark subamplitude $A_1(s, s_{12}, s_{123})$ (a), the glueball subamplitude $A_2(s, s_{12}, s_{34})$ (b) and the hadronic subamplitude $A_p(s, s_{12}, s_{34})$ (p=3, 4, 5, 6, 7) (c). The bold line corresponds to the constituent gluon contribution.

Fig. 2. Contours of integration 1, 2 in the complex plane $s_{13}$ for the functions $J_1$, $J_3$.

Fig. 3. Contours of integration 1, 2, 3 in the complex plane $s_{134}$ (a) and $s_{13}$ (b) for the function $J_2$.

Fig. 4. Contours of integration 1, 2 in the complex plane $s_{24}$ for the function $J_3$.

Fig. 1

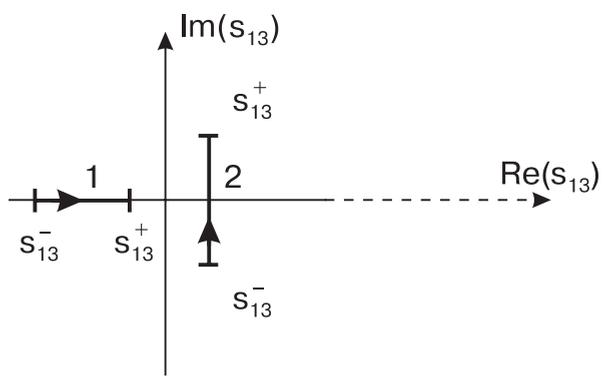

Fig. 2

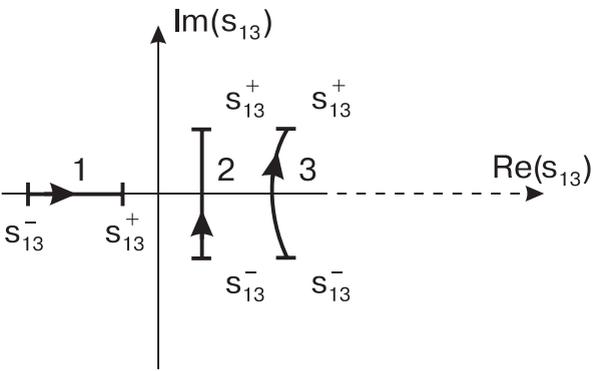

Fig. 3b

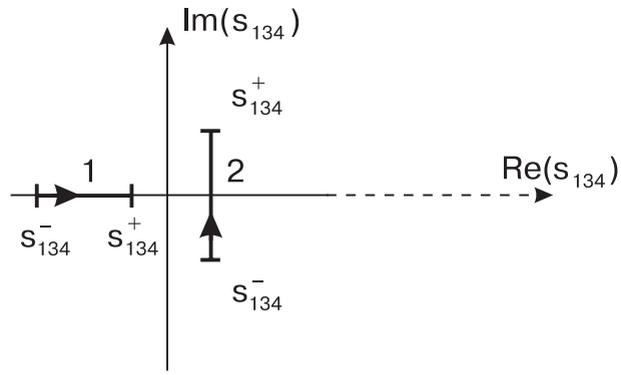

Fig. 3a

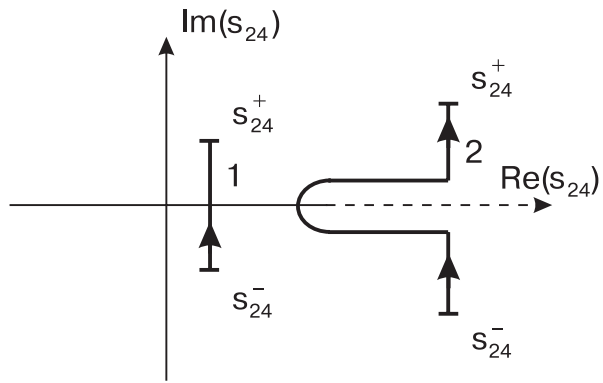

Fig. 4